\documentclass[aps,showpacs,superscriptaddress,showkeys,prl,10pt,groupedaddress,twocolumn]{revtex4-1}
\usepackage[utf8]{inputenc}
\usepackage{amsmath,amssymb,slashed}
\usepackage{float}
\usepackage{graphicx}
\usepackage{xcolor}
\usepackage{bm}
\usepackage[english]{babel}
\usepackage{comment}
\begin{document}
\title{Shadowing Neutrino Mass Hierarchy with Lorentz Invariance Violation}

\author{H. Jurkovich}
\email{jurkovic@ifi.unicamp.br}
\affiliation{Instituto de F\'isica Gleb Wataghin - Universidade Estadual de Campinas -UNICAMP, {13083-859}, Campinas SP, Brazil}
\author{Pedro Pasquini}
\email{pasquini@ifi.unicamp.br}
\affiliation{Instituto de F\'isica Gleb Wataghin - Universidade Estadual de Campinas -UNICAMP, {13083-859}, Campinas SP, Brazil}
\affiliation{Theoretical Physics Department, Fermi National Accelerator Laboratory, P. O. Box 500, Batavia, IL 60510, USA}
\author{C. P. Ferreira}
\email{cesarpf@ifi.unicamp.br}
\affiliation{Instituto de F\'isica Gleb Wataghin - Universidade Estadual de Campinas -UNICAMP, {13083-859}, Campinas SP, Brazil}

\begin{abstract}
The effects of Lorentz Invariance Violation (LIV) operators up to dimension 6 in long baseline neutrino experiments are discussed, in specific for DUNE and T2K. A phenomenological Lagrangian is proposed followed by a computation of the effective Hamiltonian of neutrino propagation in matter for mass eigenstates. It is shown that the simplest dimension 4 Lorentz violation parameter can decrease DUNE sensitivity to neutrino mass hierarchy. Also, a $\chi^2$ analysis is performed to obtain the expected long-baseline constraints to the LIV operators up to dimension 6.
\end{abstract}
\maketitle
\section{Introduction}\label{sec:intro}

In the past 40 years, the standard model has been extended in many different ways. Some extensions seek to include the standard model $SU(3)_{C}\otimes SU(2)_{L} \otimes U(1)_{Y}$ gauge group into a larger group, which is them spontaneously broken into the standard model gauge group~\cite{deBoer:1994dg}. Other extensions seek to add an additional symmetry between fermions and boson, called supersymmetry~\cite{deBoer:1994dg}. In general, both of these methods still take the Lorentz Invariance for granted, since it is well tested for most of the known particles. 
Nevertheless, models containing spontaneous Lorentz Invariance Violation (LIV), where Lorentz symmetry is spontaneously broken at Planck scale exists~\cite{Bluhm:2004ep} and might result in effects that can only be measured in very high energies not accessible to particle accelerators. 

Long baseline experiments, on the other hand, may be able to access such tiny effects since they measure phase differences of the order of 10 meV and can be very sensitive to small deviations in neutrino propagation. This can be used to observe or constraint violations of the Lorentz invariance.

\begin{figure}[hbt]
\centering
\includegraphics[scale=0.3]{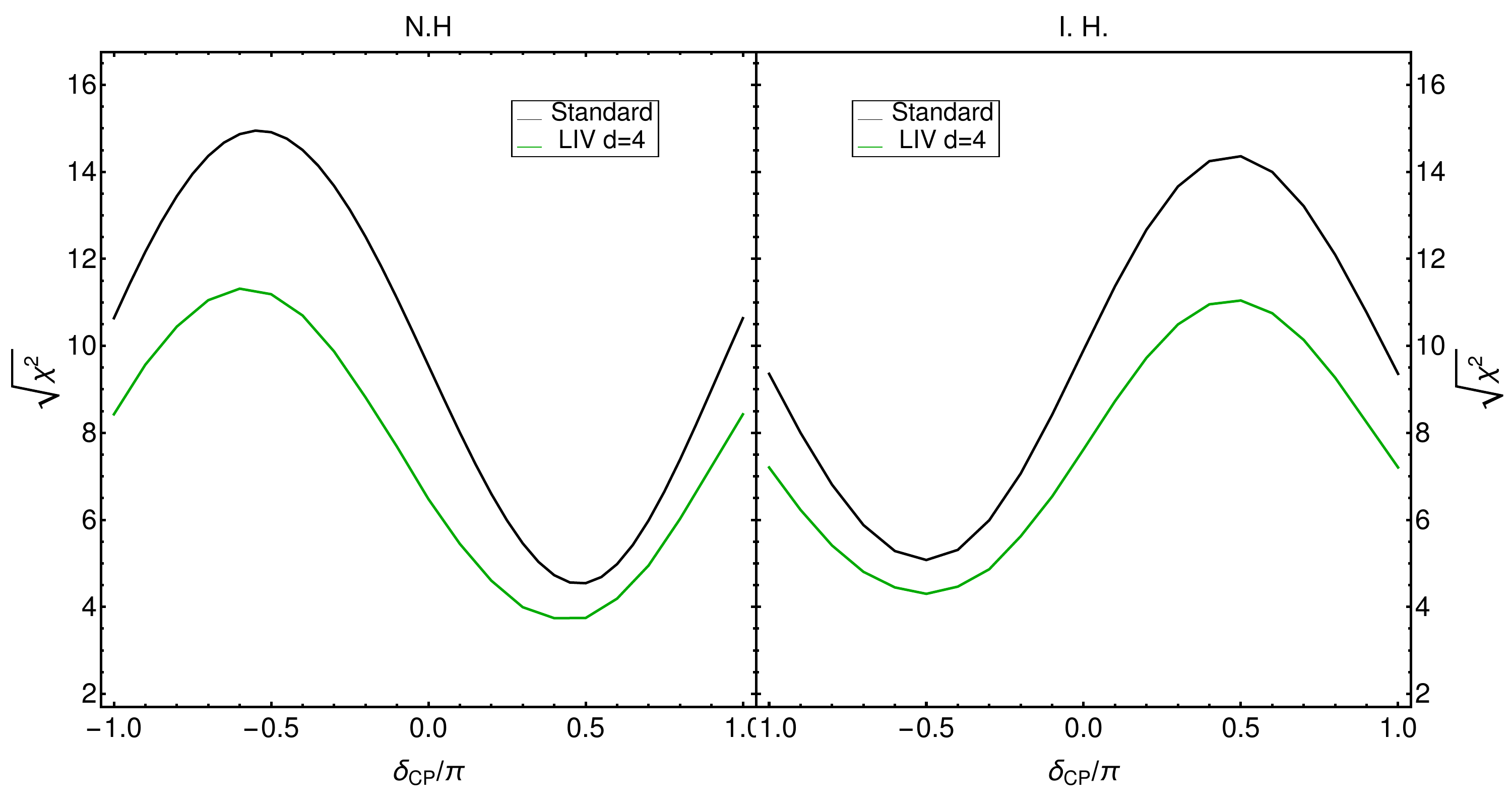}
\caption{Mass hierarchy Sensibility expected in the DUNE experiment. Black curve considers only the Standard Oscillation parameters while the Green curve considers Lorentz Violation parameter of dimension 4. {\bf Left:} Normal Hierarchy is assumed as the True Value {\bf Right:} Inverted Hierarchy is assumed as the true value.}
\label{sensitivityMH}
\end{figure}

This work is focused on the phenomenological interactions generated by theories that violate Lorentz symmetry. Using a LIV model with higher order derivatives their effect are analyzed for DUNE and T2K. In special, a closer look at mass hierarchy measurements is taken, one of the main goals of DUNE. It is shown that the dimension 4 LIV operator can decrease the sensitivity to the measurement of neutrino hierarchy. This can be observed in Fig.~\ref{sensitivityMH}, where the black curve shows the mass hierarchy sensitivity for the standard Oscillation parameters only, while the green curve includes the dimension 4 operator.

 Also, we check that operators of dimension 5 and 6 do not lower the sensitivity appreciably. 
For completeness, the expected limits for the LIV parameters are obtained for DUNE and T2K for parameters of dimension up to dimension 6.

\section{\label{sec:method}Lorentz Invariance Violating (LIV) Models}

 Lorentz violation is not a novelty in neutrino physics. It was proposed in~\cite{Coleman:1998ti} and further studied in~\cite{Cohen:2006ir,Antonelli:2018fbv}.  In this work, the focus is on models generated by LIV terms with higher order derivatives. These terms change the energy dispersion relation and as a consequence the properties of neutrino oscillation. By Considering changes on kinetics terms, the general LIV free Lagrangian is,

\begin{equation}
\mathcal{L}_{d-dim} =  i\nu^{\dagger}_{iL} \bar\sigma^{\mu} \partial_{\mu}\nu_{iL} -i^{d-3}\gamma_{i}^{j_{1}...j_{d-4}} \nu^{\dagger}_{iL} \sigma^{k} \partial_{k}\partial_{j_{1}}...\partial_{j_{d-4}}\nu_{iL},
\end{equation}

where $\gamma_{i}^{j_{1}...j_{d-4}}$ are $d-4$ tensors and $\sigma^\mu$ are the Pauli matrices. Using Euler-Lagrange equations the energy-dispersion relation in momentum space is obtained,
\begin{equation}
\sigma^{0}E\nu_{iL} = (\sigma^{k}p_{k}+\gamma_{i}^{j_{1}...j_{d-4}}\sigma^{k}p_{k}p_{j1}...p_{j(d-4)})\nu_{iL}.
\end{equation}
This equation can be written as, 

\begin{equation}
\left( \begin{array}{ccc}
E  & 0 \\
0 & E
\end{array}
\right) \nu_{iL}= (1+\overline{\gamma})
\left( \begin{array}{ccc}
p_{3} & p_{1} -ip_{2} \\
p_{1} +ip_{2} & -p_{3}
\end{array}
\right) \nu_{iL},
\end{equation}
where $\overline{\gamma}=(\gamma_{i}^{j_{1}...j_{d-4}}p_{j1}...p_{j(d-4)})$ and is an eigenvalue equation. which has an solution of the form,
\begin{equation}\label{eq:dispersion}
E^2=(1+\overline{\gamma})^2\textbf{p}^2.
\end{equation}

If massive neutrinos are considered, Eq.~\ref{eq:dispersion} is modified by 
\begin{equation}
E^2=(1+\overline{\gamma})^2\textbf{p}^2 + m^2,
\end{equation}
where we ignored terms proportional to $\overline{\gamma}m^2$, since they are small. Neutrino masses can be generated by many different theoretical approaches~\cite{King}, but none of them will change the dispersion relation. 

Notice that $\overline{\gamma}$ is momentum dependent. By Assuming the neutrino propagation direction as $x_{3}$, the expansion in the energy of the neutrino $\nu_i$ takes a simple form,
\begin{equation}
E  \approx \frac{m^2_i}{2p}+(1+\gamma^{(d)}_{i}p^{d-4})p.
\end{equation}
Which changes the usual Hamiltonian in the mass basis to
\begin{equation}\label{eq:hamil_LIV}
H\rightarrow H_0+ H_{\rm LIV},
\end{equation}
where $H_0={\rm Diag}[0 ,\Delta m^2_{21}/2E,\Delta m^2_{31}/2E]+UV(x)U^{\dagger}.$
and

\begin{equation}
H_{\rm LIV}=
\left(
\begin{array}{ccc}
0 & 0 & 0\\
0 & \Delta \gamma^{(d)}_{21}E^{d-3}& 0\\
0 & 0 & \Delta\gamma^{(d)}_{31}E^{d-3}.
\end{array}
\right) E^{d-3}.
\label{liv_matrix}
\end{equation}

A study of this kind of Hamiltonian is done in~\cite{Rossi-Torres:2013wla,Esmaili:2014ota, x}. Also, the general $d=4$ LIV scenario is considered in~\cite{x,Barenboim:2018ctx} for the DUNE experiment. However ~\cite{Barenboim:2018ctx} didn't considered its impact on standard oscillation parameters such as mass hierarchy and the constraints were obtained in the flavor basis. A map between flavor basis and mass basis for the LIV parameter is possible, but it is not trivial since in our analysis several parameters of the standard three neutrino scenario were kept free. Here, the simple case where $\Delta\gamma^{(d)}_{21}=\Delta\gamma^{(d)}_{31}=\gamma^{(d)}$ is analyzed.
\section{Analytical expression for neutrino probabilities}
\label{e-alpha}

Because the energy dependence in the Hamiltonian is going to be very important as the energy increases, it is fundamental to have an analytic expression for the neutrino probabilities in order to understand the physical consequences of this extra LIV parameter. A very similar approach as Ref.~\cite{Asano:2011nj, x} is taken. In this context, an expansion in perturbation theory is done with the assumptions that the parameters $\sin \theta_{13}$, $r_{\Delta} \equiv \frac{\Delta m^2_{21}}{\Delta m^2_{31}}$ are small. The drawbacks of this formalisms are that it does not work for low energies, near the solar resonance, but since the interest is in long-baseline experiments, this approximation should not be a problem. At first order one obtain a very similar structure to the standard three neutrinos scenario, with the following replacement:

\begin{eqnarray}\nonumber
r_{\Delta}\Delta \to r_{\Delta}\Delta+E^{d-3}\gamma^{(d)}\equiv \Delta\left(r_{\Delta}+\eta \right) \\  \Delta \to \left(\Delta+ E^{d-3}\gamma^{(d)}\right)\equiv \Delta \left( 1 +\eta\right),
\label{mudada}
\end{eqnarray} 

where $\eta = \frac{E^{d-3}\gamma^{(d)}}{\Delta}$.

\begin{figure}[hbt]
\centering
\includegraphics[scale=0.3]{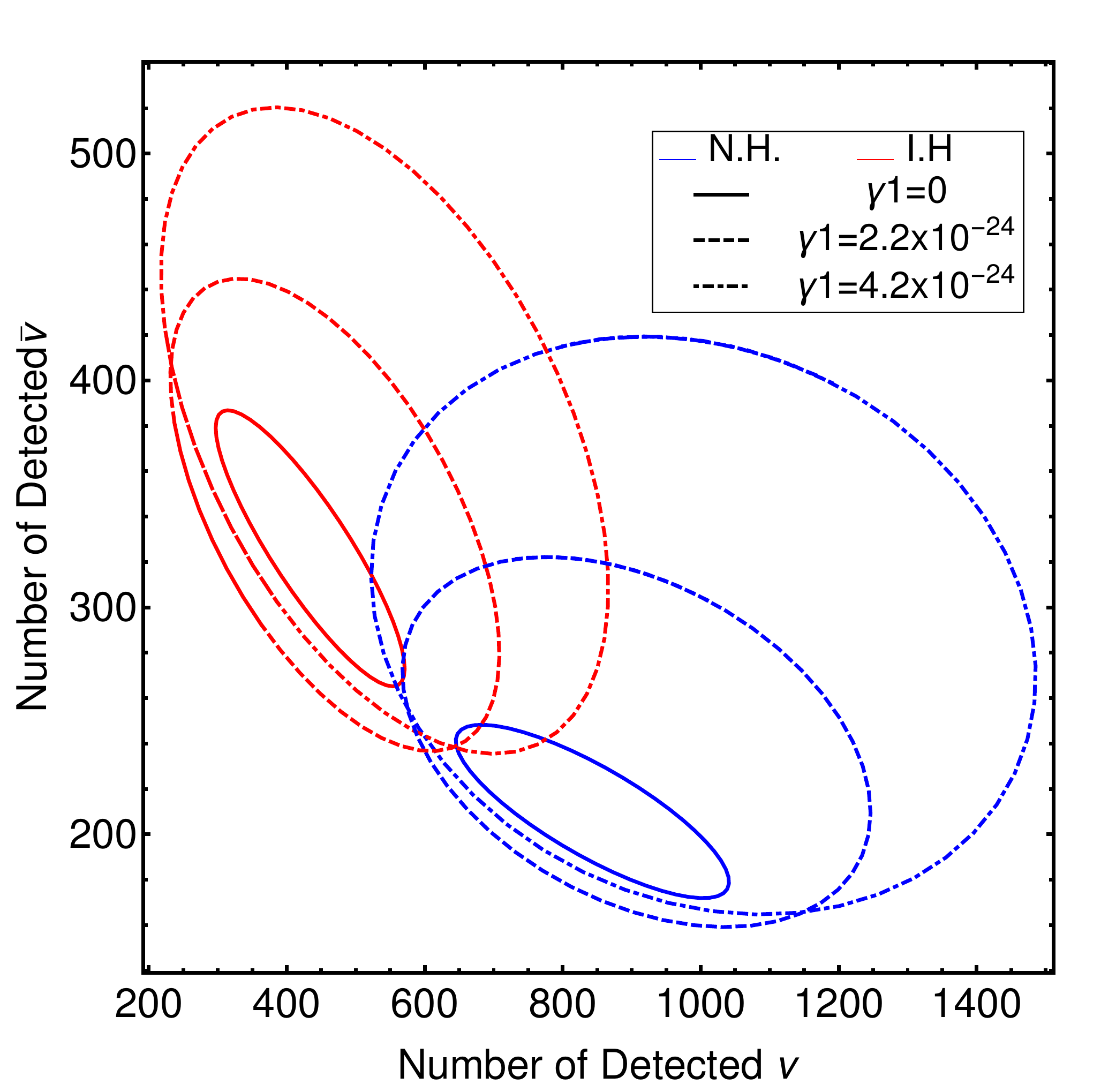}
\caption{Mass hierarchy Sensibility expected in the DUNE experiment. Thick curve considers only the Standard Oscillation parameters while the other curves considers Lorentz Violation parameter of dimension 4. {\bf Blue:} Normal Hierarchy is assumed as the True Value {\bf Red:} Inverted Hierarchy is expected as the true value.}
\label{ellipsesMH}
\end{figure}

In appendix~\ref{appendix1}, the full expression up to second order in the parameter expansion is given. The most affected probability in this expansion is the conversion probability $P_{\mu\rightarrow e}$  as can be seen in Eq.~(\ref{Pemu-1a},\ref{Pemu-3/2a},\ref{Pemu-2a}) of the Appendix. In particular, to first order, one obtains:
{\small
\begin{eqnarray}
P_{\mu \to  e }^{(1)} &=&
4 s^2_{23} s^2_{13}\left(\frac{\bm{\left(1+\eta\right)^2}}{ (1+ \bm{\eta}- r_{A})^2 }\right)
\sin^2 \frac{ (1 \bm{+\eta}- r_{A}) \Delta L }{ 2 },
\label{Pemu-1a}
\end{eqnarray} }
where, 
$s_{ij}\equiv \sin \theta_{ij}$, $r_{ \Delta } \equiv \frac{ \Delta m^2_{21} }{ \Delta m^2_{31} }$ 
$\Delta \equiv \frac{ \Delta m^2_{31} }{ 2E }$, 
$r_{A} \equiv \frac{ a }{ \Delta m^2_{31} }$. The LIV model shifts its amplitude quadratically in $\eta$ and also shift its phase from the standard case. The phase change is equivalent to the replacement of $1+r_A \to 1+\bm{\eta}+r_A$ and can lead to a decrease in the sensitivity of the mass hierarchy measurement for $d=4$. This is beautifully illustrated in the ellipses of events in Fig.~\ref{ellipsesMH}. The full line represents the number of events for normal hierarchy (N.H) in blue and inverted hierarchy (I.H.) in red by varying only the $\delta_{CP}$.

Notice that, in the standard three-neutrino scenario, those ellipses do not touch, hence DUNE can distinguish well NH from IH. However, when we turn on the $\gamma^{(4)}$ parameter, the ellipses starts to grow until they touch each other. Since the energy dependence of $\eta$ and $r_A$ are not the same, the sensitivity is not completely lost. Moreover, higher dimensions cannot properly fake the behavior of the mass squared difference, thus only $d=4$ decreases the sensitivity. This implies that there is no overlap in N.H and I.H for dimensions d=5 and 6 in a similar figure for these dimensions. The $d=4$ effect in the measurement of the mass hierarchy is illustrated in Fig~\ref{sensitivityMH}. The black line corresponds to the Standard Case while the Green Line to the LIV for $d=4$.

For completeness the limit case when the LIV parameter  $\gamma^{(d)} E^{d-3}$ is much larger than all the other parameters of the three neutrino scenario is presented, $\gamma^{(d)} E^{d-3}  \gg \Delta, r_{\Delta}\Delta$ one can get the following expressions for the probabilities:
\begin{eqnarray}\nonumber
P_{\mu\to e}=+4|U_{\mu 1}|^2 |U_{e 1}|^2  \sin^2 \left(\frac{\eta\, \Delta  L}{2}\right). \\
 P_{\mu\to \mu}=1-4|U_{\mu 1}|^2 \left( 1-|U_{\mu 1}|^2 \right)   \sin^2 \left(\frac{\eta \, \Delta L}{2} \right).
\label{bigneg}
\end{eqnarray}

In this limit, the neutrino probabilities are even functions of $\gamma^{(d)}$, meaning that $\gamma^{(d)}$ and $-\gamma^{(d)}$ gives the same results. Also, they are equal for neutrinos and anti-neutrinos.

\begin{figure*}[hbt]
\centering
\includegraphics[scale=0.2]{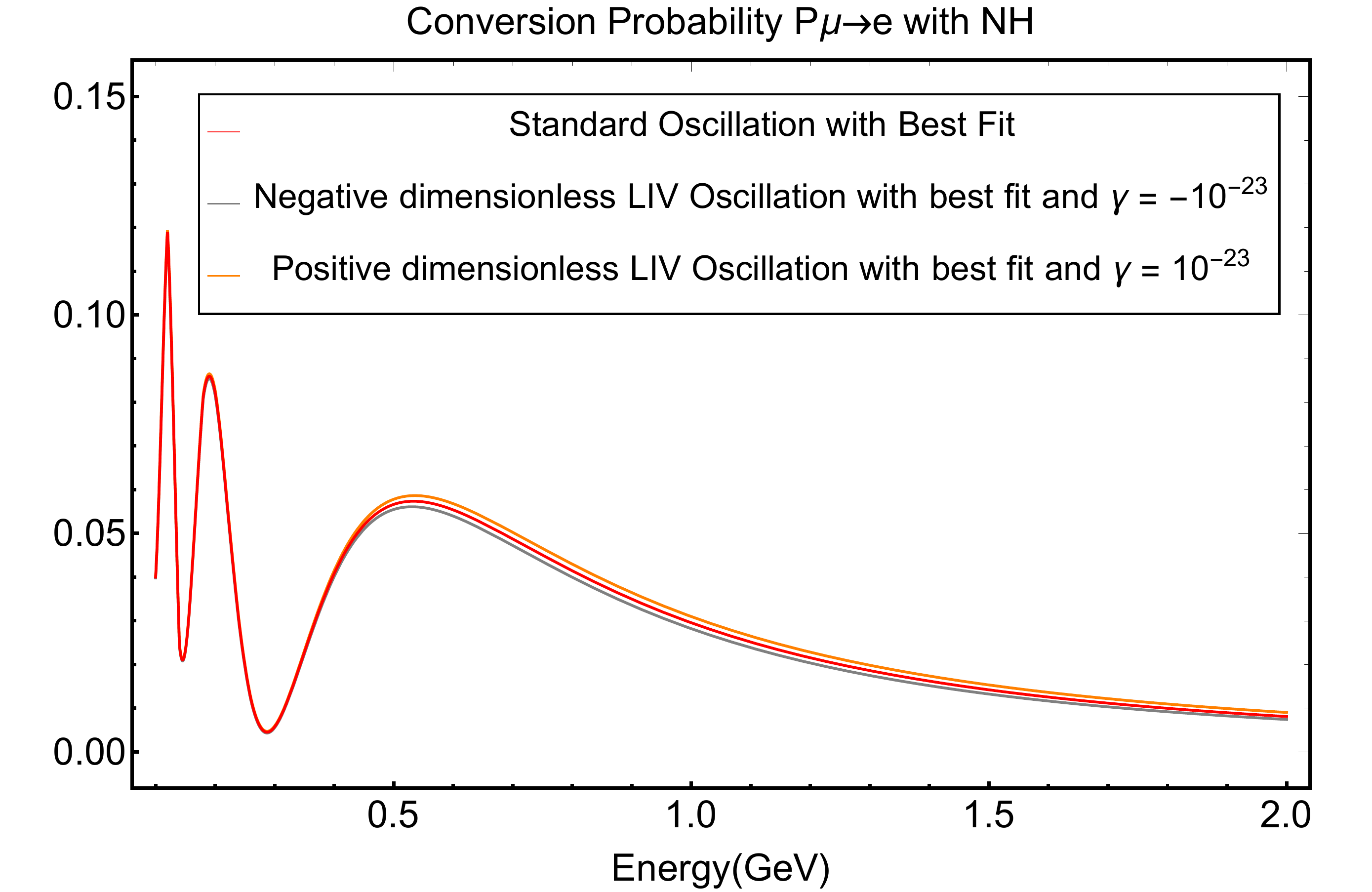}
\includegraphics[scale=0.2]{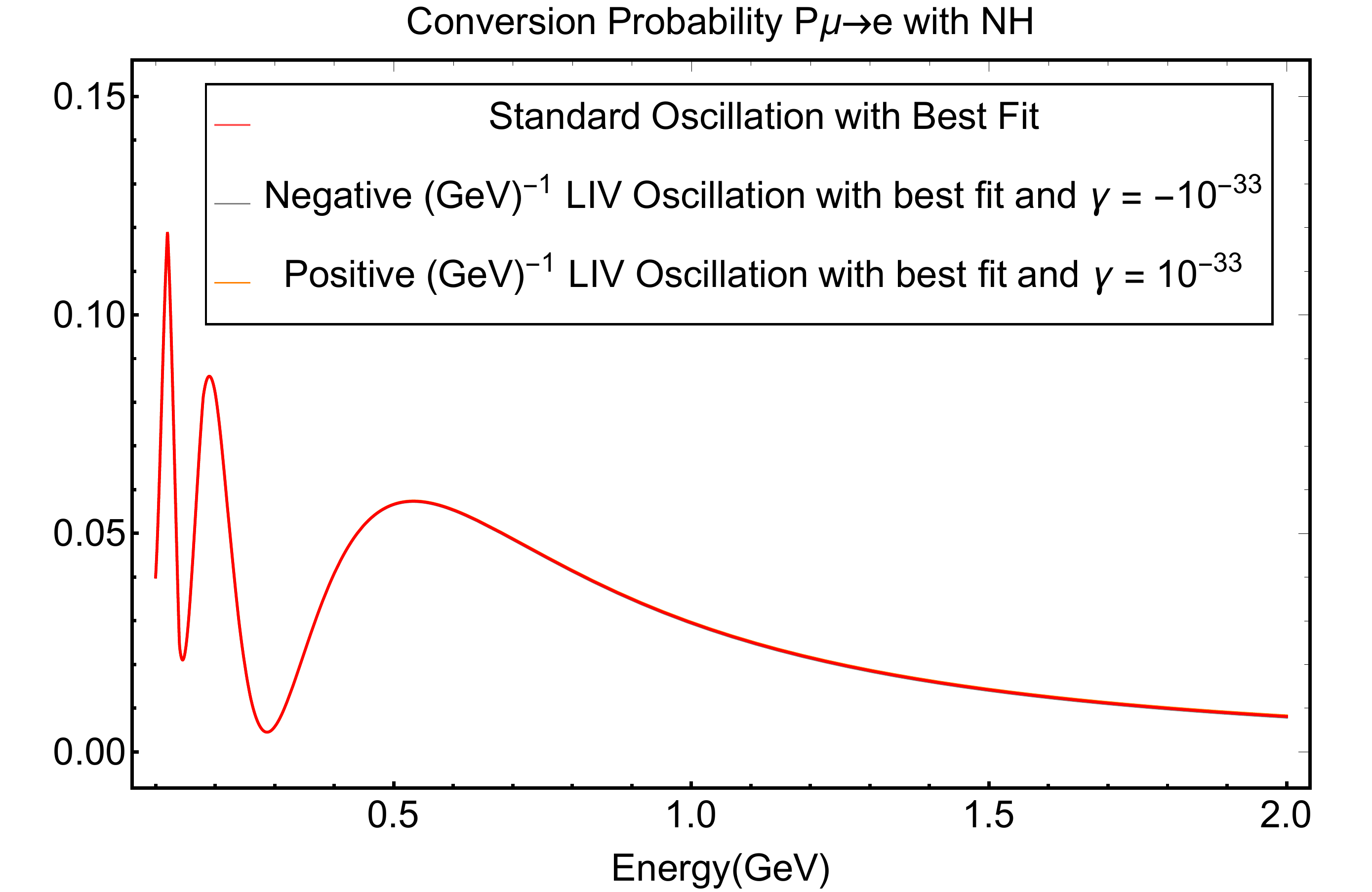}
\includegraphics[scale=0.2]{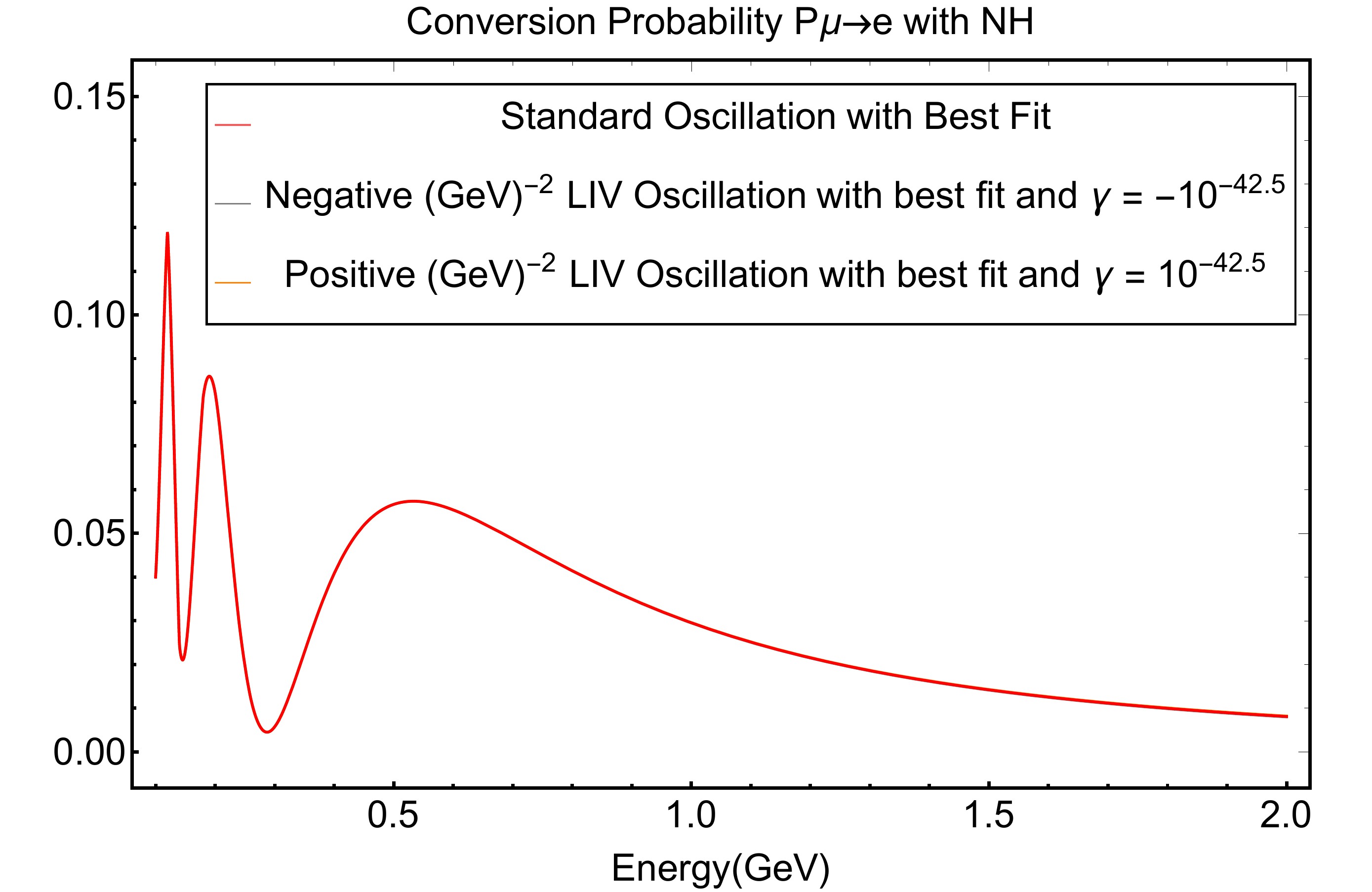}\\
\includegraphics[scale=0.2]{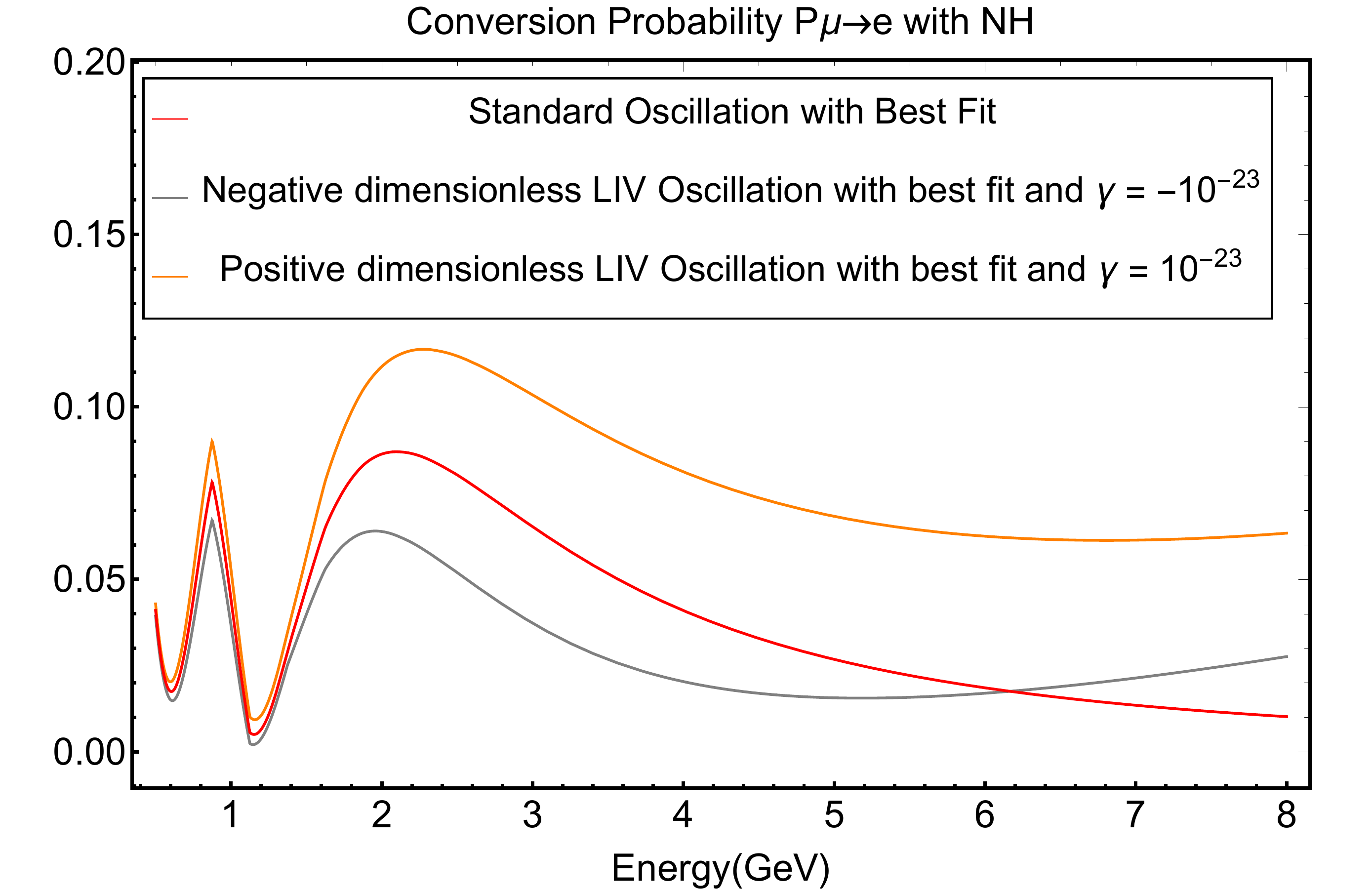}
\includegraphics[scale=0.2]{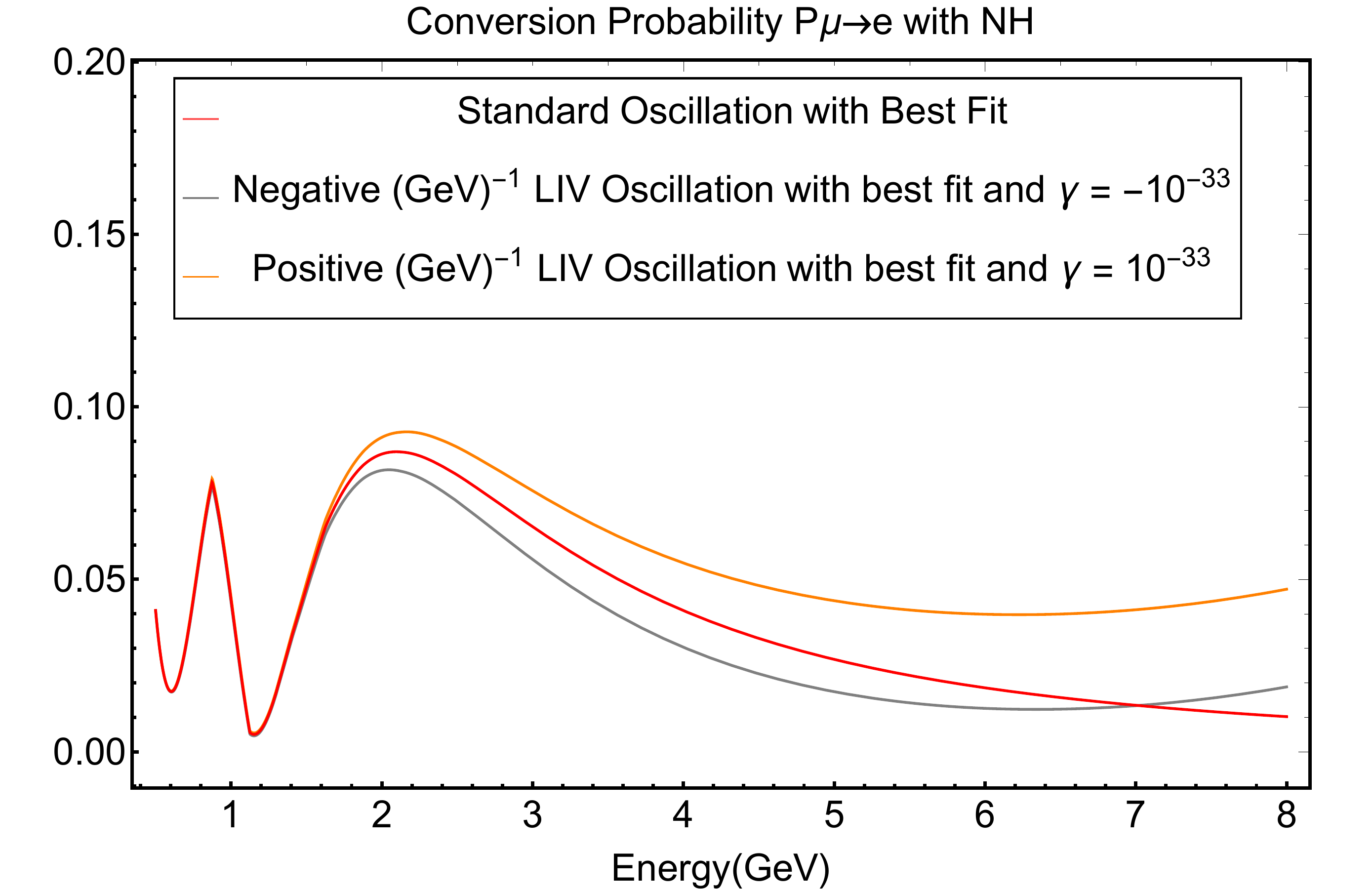}
\includegraphics[scale=0.2]{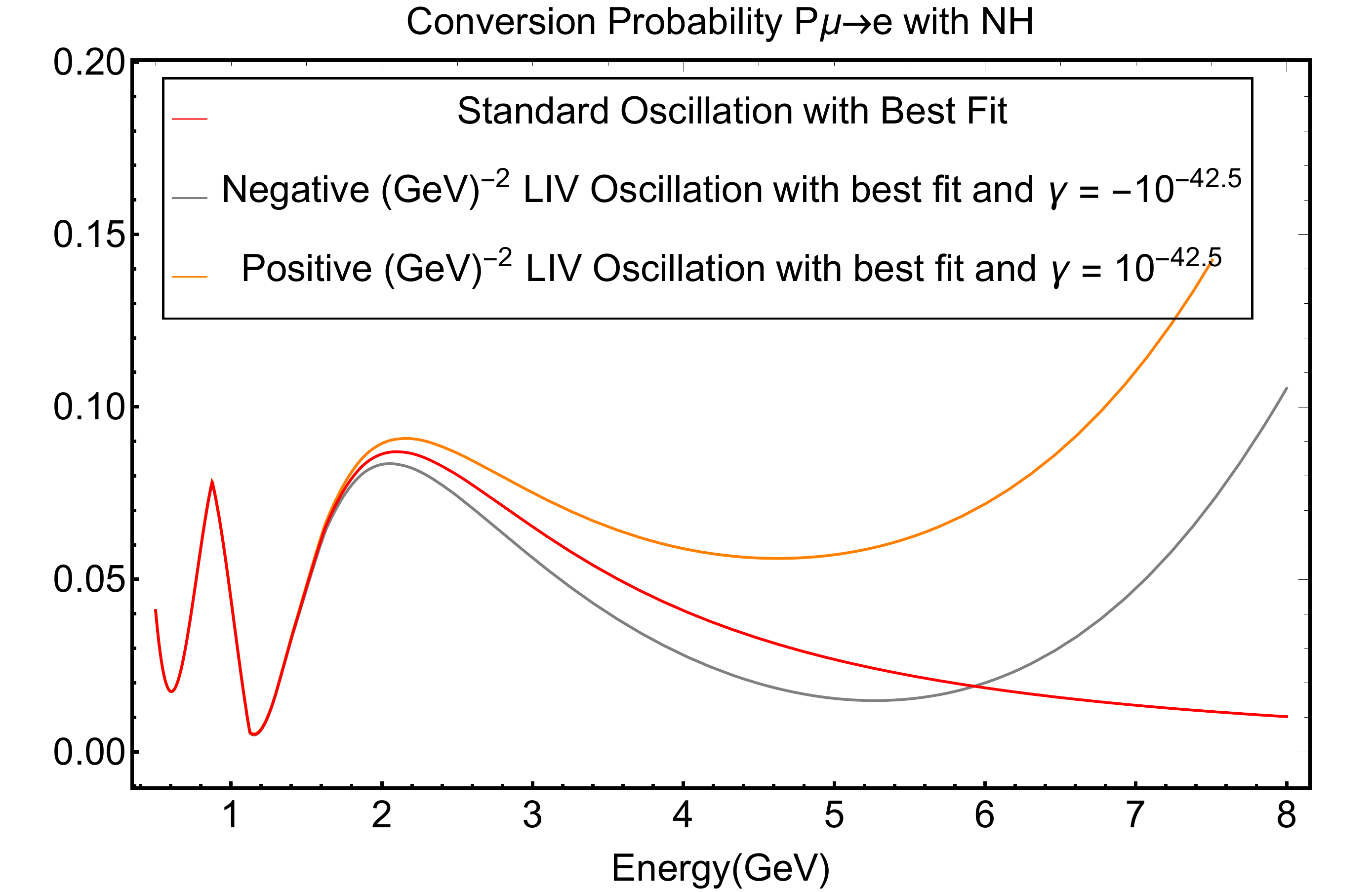}
\caption{Conversion probabilities for {\bf TOP:} T2K and {\bf BOTTOM:} DUNE using operators of mass dimension 4, 5 and 6 from the left to the right. In the boxes you can find the values for the $\gamma^{(d)} = \gamma$ parameters for each curve. }
\label{conv1}
\end{figure*}

The exact probabilities are plotted for DUNE and T2K in Fig.~\ref{conv1}. The conversion of muon neutrinos to electron neutrinos for a fixed baseline of $L=1300$~km for DUNE and $L=295$~km for T2K, is done by numerically solving the differential equation shown in Eq.~\ref{liv_matrix}. The initial conditions are $\Psi_\mu(0)=(\Psi_{\mu e}~\Psi_{\mu\mu}~\Psi_{\mu\tau})=(0~1~0)$. 

The conversion probabilities for DUNE and T2K are obtained using Table.~\ref{tab_osc_par} as the best fit results for the standard three neutrino scenario:

\begin{table}[ht]
\centering
\begin{tabular}{cccccc}
\hline
&$\Delta m^2_{21}$~(eV$^2$) & $\Delta m^2_{31}$~(eV$^2$) & $\sin^2\theta_{21}$ & $\sin^2\theta_{23}$ & $\sin^2\theta_{13}$\\
\hline
NH & $7.56\times 10^{-5}$ & $2.55\times 10^{-3}$ & 0.321  & 0.430 & 0.0216 \\
\hline
IH & $7.56\times 10^{-5}$ & $2.49\times 10^{-3}$ & 0.321 & 0.596 & 0.0214\\
\hline
\end{tabular}
\caption{Values of the best-fit oscillation parameters extracted from the Reference~\cite{Escrihuela:2016ube}.}
\label{tab_osc_par}
\end{table}

As one can see in Fig.~\ref{conv1}, the behavior described by the analytical expressions are observed. First, is analyzed the case where $d=4$, Eq.~\ref{Pemu-1a} shows that the amplitude growth as $(1+\eta)^2$ and since $\eta =
\frac{\gamma^{(d)} E^{d-3}}{\Delta}$ depends on the energy it increases the amplitude as the energy increases. Since the argument of Eq.~\ref{Pemu-1a} also depends on $\eta$ it also shifts the position of the peak. This behavior is more dramatic when $d=5$ and $d=6$. Here, even when a value for $\gamma^{(d)}$ is chosen that does not increase the amplitude at lower energies, as the energy increases and this term becomes bigger with a power of $E^{d-4}$ and the probability deviates a lot from the standard three-neutrino scenario. Moreover, when the energy is about 8~GeV the negative LIV term starts to increase the probability, the reason for this behavior is that the LIV term becomes so much bigger in the arguments of the oscillatory terms than the usual terms of the three neutrino scenario that it dominates them completely (see Eq.~\ref{bigneg}).

Using the same parameters for $\gamma^{(d)}$ that was used for DUNE one can notice that in Fig.~\ref{conv1} the effects of LIV are small in T2K in comparison to the DUNE results. This is a consequence of the different energy range and baseline between both experiments. T2K's energy/baseline are smaller compared to DUNE. This is illustrated in Eq.~\ref{Pemu-1a}. The amplitude of the conversion probability depends polynomially on the parameter $\eta\sim E^{d-3}$. Therefore, higher energy beams are more sensitive to higher dimensional Lorentz violation.
 
\section{\label{sec:method1}Experimental Simulation}
For completeness, the expected bound reachable by DUNE and T2K experiments for each of the LIV parameters with $d=4,5,6$ is presented. In order to simulate the new physics, a modified version of Globes~\cite{Huber:2007ji} probability engine was used in order to construct the Hamiltonian of Eq.~\ref{eq:hamil_LIV}. The assumed experimental configuration is described below.
\begin{itemize}
\item[1.] {\bf T2K:} Tokai to Kamiokande (T2K) experiment~\cite{Abe:2014tzr} is a long baseline neutrino facility localized in Japan. The beamline containing mostly $\mu$-neutrino (anti-neutrino) leaves Tokai arriving at Kamioka mine 295 km away in an off-axis (by $2.5^\circ$) water Cherenkov tank of 50~ kt (fiducial volume 22.5 Kt) of mass: The Super-Kamiokande detector~\cite{Abe:2016nxk}. Our simulation assumes a neutrino energy flux around 0.6 GeV and a total exposure of $7.8\times 10^{21}$ protons on target (POT), with a proportion of 50\%/50\% of neutrino/anti-neutrino mode. The expected normalization error can be as low as 5\% (10\%) for the signal (background).

\item[2.] The Deep Underground Neutrino Experiment (DUNE) consists of a 40 kt (fiducial) liquid Argon neutrino detector located at Sanford Underground Research Facility which is 1300 km away of its neutrino beam source at Fermilab. Neutrinos are created by pion and kaon decays-in-fight. Which provides an energy neutrino flux at around 2.5 GeV and can reach a 1.47$\times 10^{21}$ POT/year exposure. A detailed description of the experiment can be found in~\cite{Acciarri:2015uup,Alion:2016uaj}. Here we assume a 3.5 years run for each neutrino and anti-neutrino mode and 4\% (10\%) of signal (background) normalization error.
\end{itemize}

\begin{figure*}[hbt]
\centering
\includegraphics[scale=0.5]{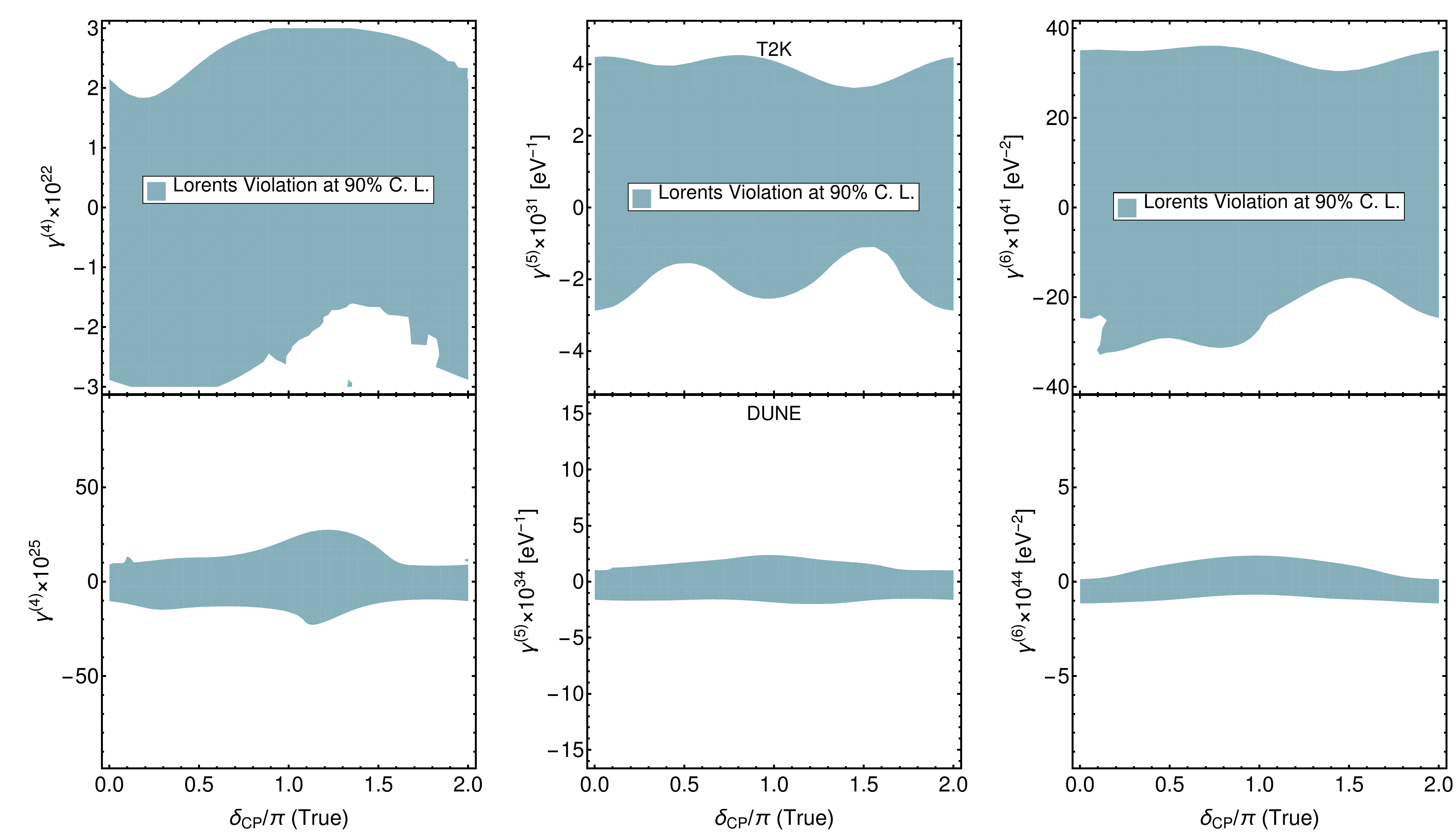}
\caption{Exclusion region for the LIV parameter $\gamma^{(d)}$ of mass dimension 4, 5 and 6 versus $\delta$ at 90\% C.L. Assuming N.H. {\bf Top:} T2K and {\bf Bottom:} DUNE.}
\label{margt2k}
\end{figure*}

The sensitivity analysis was performed assuming a $\chi^2$ distribution divided into two factors, 
\begin{equation}
\chi^2=\chi^2_P+\chi^2_{sys},
\end{equation}
$\chi^2_P$ corresponds to Pearson's statistical distribution for the number of events,
\begin{equation}
\chi^2=\sum_i \left(\frac{N_i^O-(1-a)N_i^s-(1-b)N_i^b}{\sqrt{N_i^O}}\right)^2,
\end{equation}
where $N_i^O$ is the expected true number of neutrinos detected in bin $i$, $N_i^s$ is the test number of signal neutrinos and $N_i^b$ is the test number of background neutrinos. The $a,b$ are auxiliary parameters that implement the normalization error through: 
\begin{equation}
\chi^2_{sys}=\left(\frac{a}{\sigma_a}\right)^2+\left(\frac{b}{\sigma_b}\right)^2,
\end{equation}
with $\sigma_a$ ($\sigma_b$) the signal (background) normalization error. In order to extract the sensitivity to LIV parameters we minimize over all test oscillation, $a$ and $b$ and standard parameters for each fixed true value of $\delta_{\rm CP}$ and test value of $\gamma^{(d)}_i$. All true LIV parameters set to zero. The result can be found on Fig.~(\ref{margt2k}) for T2K and DUNE on at 90\% of C. L. Notice that $\delta_{\rm CP}$ (true) has only a small impact on the sensitivity of the LIV parameters and that DUNE is much more sensitive (three orders of magnitude) to the LIV parameters as expected from Fig.~\ref{conv1}.

The conservative DUNE constraints in the LIV parameters for operators of dimension 4, 5 and 6 can be found on Table~\ref{tab_osc_par}.

\begin{table}[ht]
\centering
\small
\begin{tabular}{c|cccc}
\hline
& $|\gamma^{(4)}|$ & $|\gamma^{(5)}|$ $ GeV^{-1}$ & $|\gamma^{(6)}|$ $GeV^{-2}$  \\
\hline
\hline
T2K & $4.1\times10^{-21}$ & $4.6\times10^{-31}$ & $3.7\times10^{-41}$\\
DUNE & $8\times10^{-24}$ & $6.7\times10^{-34}$ & $1.2\times10^{-44}$\\
\hline
\end{tabular}
\caption{constrains in LIV operators of mass dimension 4, 5 e 6 at 2$\sigma$ with $\delta_{CP}$(true)=$1.4\pi$.}
\label{tab_osc_par}
\end{table}

\section{Conclusion}
Long-baseline neutrino experiments can probe Lorentz Invariance Violation due to its incredible precision in measuring phase differences in neutrino propagation. Thus, the potential of current and future neutrino experiments such as DUNE and T2K to constraint LIV parameters of dimension $d=4,5,6$ is shown. In particular, DUNE can be affected by LIV parameters of dimension $d=4$ and lose sensitivity to mass hierarchy, one of its main goals to achieve.

\onecolumngrid

\section{Appendix} \label{appendix1}

 Applying perturbation theory to this LIV model~\cite{x}, one obtain a very similar structure to the standard three neutrinos scenario, with the following replacement:

\begin{eqnarray}\nonumber
r_{\Delta}\Delta \to r_{\Delta}\Delta+E^{d-3}\gamma^{(d)}\equiv \Delta\left(r_{\Delta}+\eta \right). \\ \Delta \to \left(\Delta+ E^{d-3}\gamma^{(d)}\right)\equiv \Delta \left( 1 +\eta\right),
\label{mudada}
\end{eqnarray} 

where $\eta = \frac{E^{d-3}\gamma^{(d)}}{\Delta}$.
Then the LIV modified probabilities becomes:

\begin{eqnarray}
\left(P_{\mu \to  e }^{(1)}\right)^{\rm VLI} &=&
4 s^2_{23} s^2_{13}\left(\frac{\bm{\left(1+\eta\right)^2}}{ (1+ \bm{\eta}- r_{A})^2 }\right)
\sin^2 \frac{ (1 \bm{+\eta}- r_{A}) \Delta L }{ 2 },
\label{Pemu-1a} \\
\left(  P_{\mu \to e}^{(3/2)}\right)^{\rm VLI} &=&
8 J_{r} \left(\frac{ r_{\Delta}+\bm{\eta}}{r_{A} (1 +\bm{\eta}- r_{A})} \right)\left(1+\bm{\eta}\right)
\cos \left( -\delta - \frac{ \Delta\left(1+ \bm{\eta}\right) L }{ 2 } \right)
\sin \frac{ r_{A} \Delta L }{ 2 }
\sin \frac{  (1+\bm{\eta}- r_{A}) \Delta L }{ 2 },
\label{Pemu-3/2a} \\
 \left( P_{\mu \to  e}^{(2)} \right)^{\rm VLI} &=& 
4 c^2_{23} c^2_{12} s^2_{12} 
\left( \frac{ r_{\Delta}+\bm{\eta}}{ r_{A} } \right)^2 
\sin^2 \frac{ r_{A} \Delta L }{ 2 } 
\nonumber \\
&-& 
4 s^2_{23} \left[ 
s^4_{13}\left( \frac{ \left(5 \bm{\eta}^2 - 2 \bm{\eta} (-3 + r_A) + (1 + r_A)^2\right) \bm{(1+\eta)} }{ (1+\bm{\eta} - r_{A})^4 } \right)
- 2 s^2_{12} s^2_{13} \frac{( r_{\Delta} +\bm{\eta}) 
\left(r_{A}-\bm{\eta  s_{12}^2}\right) }{(1+\bm{\eta}- r_{A})^3 }
\right] 
\sin^2 \frac{ ( 1+\bm{\eta} - r_{A}) \Delta L }{ 2 } 
\nonumber \\
&+& 
2 s^2_{23} \left[ 
2 s^4_{13} \frac{\left( r_{A}+ \bm{\eta+\eta^2} \right) (1+\bm{\eta})}{ (1 +\bm{\eta} - r_{A})^3 } 
- s^2_{12} s^2_{13} \frac{\left( r_{\Delta}+\bm{\eta}\right) 
\left(1+ \bm{\eta c_{12}^2} \right)}{ (1+\bm{\eta} - r_{A})^2 } 
\right] 
(\Delta L ) \sin (1+\bm{\eta} - r_{A}) \Delta L.  \nonumber \\
&+& 
8 \bm{\eta} s^2_{23}s^2_{12}c_{12}^2  s^2_{13}  \left[ 
 \frac{ (r_{\Delta} +\bm{\eta})} 
  {(1+\bm{\eta}- r_{A})^2 } 
\right]  \sin \left(\frac{(1+\bm{\eta} - r_{A}) \Delta L}{2}\right) \cos \left(\frac{(1+\bm{\eta} ) \Delta L}{2}\right)
 \sin \left(\frac{ r_{A} \Delta L}{2}\right)\nonumber \\
 &+& 
 4 \bm{\eta} s^2_{23}s^4_{12}  s^2_{13}  \left[ 
 \frac{( r_{\Delta} +\bm{\eta})(1\bm{\eta})} 
  {(1+\bm{\eta}- r_{A})^2 } 
\right]  \sin^2 \left(\frac{(1+\bm{\eta} - r_{A}) \Delta L}{2}\right),
\label{Pemu-2a}
\end{eqnarray}

$s_{12}\equiv \sin \theta_{12},s_{13}\equiv \sin \theta_{13},s_{23}\equiv \sin \theta_{23}$,  $r_{ \Delta } \equiv \frac{ \Delta m^2_{21} }{ \Delta m^2_{31} }$ 
$\Delta \equiv \frac{ \Delta m^2_{31} }{ 2E }$, 
$r_{A} \equiv \frac{ a }{ \Delta m^2_{31} }$ and the reduced Jarlskog coefficient
$J_{r} \equiv c_{12} s_{12} c_{23}  s_{23} s_{13}$. The superscribe in $P_{\mu e}^{(i)}$ indicates the {\it i}-th  order of expansion for the $r_{\Delta}$ and $s_{13}$ parameters. As expected, in the limit $\eta=0$ the usual three neutrino scenario formulas are recovered.

Applying the perturbation theory, for the muon survival probability, in the  LIV scenario, one can obtain:

\begin{eqnarray}
 \left(P_{\mu \to  \mu}^{(0)}\right)^{\rm VLI} &=&
1 - 4 c^2_{23}  s^2_{23} \sin^2 \left( \frac{ \Delta \left(1+\bm{\eta}\right)  L }{ 2 } \right),
\label{Pmumu-0VLI}\\
\left(P_{\mu\to  \mu}^{(1)}\right)^{\rm VLI}  &=&
 -4 s^4_{23} s^2_{13}\left( \frac{1+\bm{\eta} }{ (1+\bm{\eta} - r_{A})^2 }\right)
\sin^2 \frac{ (1+\bm{\eta}- r_{A}) \Delta L }{ 2 }
\nonumber \\
&-&
2 c^2_{23}  s^2_{23}
\left[
s^2_{13} 
\left( \frac{\bm{\eta}(1+\bm{\eta})+ r_{A} }{ 1+\bm{\eta} - r_{A} }\right) - c^2_{12}( r_{\Delta}+\bm{\eta})
\right]
( \Delta L ) \sin \Delta (1+\bm{\eta}) L
\nonumber \\
&+&
4 c^2_{23}  s^2_{23} s^2_{13}\left( \frac{(1+\bm{\eta})^2}{ (1+\bm{\eta}- r_{A})^2 }\right)
\sin \frac{ (1 +\bm{\eta}+ r_{A}) \Delta L }{ 2 } \sin \frac{ (1+\bm{\eta} - r_{A}) \Delta L }{ 2 },
\label{Pmumu-1VLI}
\end{eqnarray}
again, in the limit of $\eta=0$, the three standard neutrino scenario survival probability is recovered.

\end{document}